\documentclass[english,preprint,aps,prd,showpacs,superscriptaddress,nofootinbib,tightenlines]{revtex4}

\usepackage{graphicx}
\usepackage{booktabs}
\usepackage{graphicx}
\usepackage{epstopdf}
\usepackage{amssymb,bm,mathrsfs,bbm,amscd}
\usepackage[tbtags]{amsmath}
\usepackage{lastpage}
\usepackage{ulem}
\usepackage{color}
\usepackage{amsfonts}
\usepackage{mathrsfs}
\usepackage{amsmath}
\usepackage{amssymb}
\usepackage{bm}
\usepackage{bbm}
\usepackage{epsfig}
\usepackage{multirow}
\usepackage{float}
\usepackage{array}
\makeatletter
\makeatletter
\def\hlinew#1{%
  \noalign{\ifnum0=`}\fi\hrule \@height #1 \futurelet
   \reserved@a\@xhline}
\usepackage{array}
\newcommand{\Ga}{\Gamma}

\newcommand{\rr}{\rightarrow}
\newcommand{\fr}{\dfrac}
\newcommand{\la}{\langle}
\newcommand{\ra}{\rangle}
\newcommand{\PreserveBackslash}[1]{\let\temp=\\#1\let\\=\temp}
\newcolumntype{C}[1]{>{\PreserveBackslash\centering}p{#1}}
\newcolumntype{R}[1]{>{\PreserveBackslash\raggedleft}p{#1}}
\newcolumntype{L}[1]{>{\PreserveBackslash\raggedright}p{#1}}

\newcommand{\nn}{\nonumber}

\newcommand{\beq}{\begin{equation}}
\newcommand{\eeq}{\end{equation}}
\newcommand{\bqa}{\begin{eqnarray}}
\newcommand{\eqa}{\end{eqnarray}}

\def\OMIT#1{}

\makeatother

\usepackage{babel}

\begin{document}

\title{Relativistic correction to gluon fragmentation function into pseudoscalar quarkonium}

\author{Xiangrui Gao\footnote{gaoxr@ihep.ac.cn}}
\affiliation{Institute of High Energy Physics and Theoretical Physics Center for
Science Facilities, Chinese Academy of Sciences, Beijing 100049, China\vspace{0.2cm}}

\author{Yu Jia\footnote{jiay@ihep.ac.cn}}
\affiliation{Institute of High Energy Physics and Theoretical Physics Center for
Science Facilities, Chinese Academy of Sciences, Beijing 100049, China}
\affiliation{Center for High Energy Physics, Peking University, Beijing 100871, China\vspace{0.2cm}}

\author{LiuJi Li\footnote{liuji.li@icloud.com}}
\affiliation{Institute of High Energy Physics and Theoretical Physics Center for
Science Facilities, Chinese Academy of Sciences, Beijing 100049, China\vspace{0.2cm}}

\author{Xiaonu Xiong\footnote{xiaonu.xiong@pv.infn.it}}
\affiliation{Istituto Nazionale di Fisica Nucleare, Sezione di Pavia, Pavia, 27100, Italy}

\date{\today}

\begin{abstract}
Inspired by the recent measurements of the $\eta_c$ meson production at LHC experiment,
we investigate the relativistic correction effects for partons to fragment into $\eta_c$,  
which constitute the crucial nonperturbative elements to account for 
$\eta_c$ production at high $p_T$.
Employing several distinct methods, we calculate the leading relativistic
correction to the $g\to\eta_c$ fragmentation function in the NRQCD factorization framework,
as well as verify the existing result on relativistic correction to the $c\to \eta_c$ fragmentation function.
We also study the evolution behavior of these fragmentation functions
with the aid of the DGLAP equation.
\end{abstract}


\footnotetext[0]{\hspace*{-3mm}\raisebox{0.3ex}{$\scriptstyle\copyright$}2016
Chinese Physical Society and the Institute of High Energy Physics
of the Chinese Academy of Sciences and the Institute
of Modern Physics of the Chinese Academy of Sciences and IOP Publishing Ltd}%

\maketitle

\section{Introduction}

Heavy quarkonium production and polarization in various collider experiments
has long been a fantastic topic in QCD, which has triggered intensive
experimental and theoretical investigation in the past several decades 
(For a recent review, see \cite{Brambilla:2010cs}).

Thus far, the modern theoretical method to tackle heavy quarkonium (exemplified by $J/\psi(\psi^\prime)$ and $\Upsilon$)
production and decay is represented by the effective-field-theory approach dubbed
nonrelativistic QCD (NRQCD) factorization~\cite{Bodwin:1994jh}.
For the production of a charmonium state $H$ from the colliding beams composed of the particles of type $A$ and $B$,
NRQCD factorization allows one to, schematically, 
express the corresponding production rate as
%------------------------------------------
\beq
%------------------------------------------
d\sigma[A+B\to H+X] = \sum_n d\hat{\sigma}_n [A+B\to c\bar{c}(n)+X] \langle {\cal O}^{H}_n \rangle,
%------------------------------------------
\label{NRQCD:factorization:prod}
%------------------------------------------
\eeq
%------------------------------------------
where $n$ signifies the color/angular-momentum quantum number of the 
$c\bar{c}$ pair produced in the hard scattering.
In (\ref{NRQCD:factorization:prod}), $d\hat{\sigma}_n$ are the perturbatively-calculable short-distance coefficients, and
$\langle {\cal O}^{H}_n \rangle$ represent the nonperturbative, yet universal vacuum expectation values of the NRQCD production operators that are sensitive to $H$ and $n$.
The power series in (\ref{NRQCD:factorization:prod})
is governed by the expansion in the characteristic velocity of the $c$($\bar{c}$) quark inside a charmonium, $v$,
(the nonrelativistic nature of quarkonium implies that $v\ll 1$), since each nonperturbative NRQCD production matrix element 
possesses definite power counting in $v$.

As an important theoretical progress in the past decade,
various short-distance coefficients $d\hat{\sigma}_n$ relevant to $J/\psi$ production in virtually
all the commissioning collider programs,
have been gradually available to next-to-leading order (NLO) accuracy in strong coupling constant,
for both color-singlet and octet channels~\cite{Brambilla:2010cs}.
By confronting these NLO-accuracy NRQCD predictions with the various measurements
conducted at $B$ factorties, HERA, Tevatron and LHC, one observes some satisfactory agreement in some cases,
but also see alarming discrepancies in other cases, notoriously for $J/\psi$ polarization in hadroproduction~\cite{Brambilla:2010cs}.
Unfortunately, the current computational technique hinders our capability to further address the
next-to-next-to-leading order (NNLO) perturbative corrections, therefore,
the viability of NRQCD approach still awaits a sharper and 
more critical examination.

Very recently, for the first time, the LHCb collaboration has measured the differential production rate of the pseudoscalar
charmonium state, $\eta_c$ in the range $p_T(\eta_c)> 6.5\;\text{GeV}$, tagged via the decay channel
$\eta_c \to p \bar{p}$~\cite{Aaij:2014bga}.
This is an important supplement to our knowledge on charmonium production, since $\eta_c$ production is an even more
ideal testing-bed for NRQCD than $J/\psi$, owing to its simplicity as a spin-zero meson.
We note that, very recently, the NLO perturbative corrections to $\eta_c$ hadroproduction
have been investigated in NRQCD factorization approach~\cite{Butenschoen:2014dra,Han:2014jya,Zhang:2014ybe}.

Besides NRQCD factorization, there exists another famous first-principle approach to tackle
inclusive single hadron production, the so-called perturbative QCD (collinear) factorization,
whose applicability is not confined to merely heavy quarkonium.
According to the collinear factorization theorem~\cite{Collins:1989gx},
at sufficiently high $p_T$, the inclusive production rate of a specific hadron
$H$ is dominated by the following fragmentation mechanism:
%------------------------------------------
\bqa
%------------------------------------------
d\sigma[A+B\to H(p_T)+X] = \sum_{i}
d\hat{\sigma}[A+B\to i(p_T/z)+X]\otimes D_{i\rightarrow H}(z,\mu) + {\mathcal O}(1/p_T^2).
%------------------------------------------
\label{pQCD:factorization:large:pt}
%------------------------------------------
\eqa
%------------------------------------------
where $i$ stands for a QCD parton (quark or gluon), and $z$ is 
the light-cone momentum fraction carried by $H$ with respect to the parent parton.
$d \hat{\sigma} [A+B\to i +X]$ is the perturbatively-calculable partonic hard cross section,
$D_{i\to H}$ is the nonpertubative yet universal fragmentation function, characterizing the probability distribution 
for the parton $i$ to hadronize into $H$ carrying the momentum fraction $z$.
$\otimes$ indicates that the hard partonic cross section
ought to be convoluted with the corresponding fragmentation function over $z$.

The fragmentation functions such as a gluon fragmenting into $\pi$ and $p$ 
are genuinely nonperturbative objects, which so far can only
be extracted from experiments~\cite{Agashe:2014kda}.
In contrast, the situation becomes greatly simplified if $H$ is a heavy quarkonium.
In this case, the fragmentation function $D_{i\rightarrow H}(z,\mu)$ contains several 
distinct energy scales: heavy quark mass $m$, typical
three-momentum of quark $mv$, and even smaller scales such as  $mv^2$ and $\Lambda_{\rm QCD}$. 
Owing to the fact $m\gg \Lambda_{\rm QCD}$,
it is conceivable that the hard scale $m$ should be explicitly factored out 
from $D_{i\rightarrow H}(z,\mu)$. As a matter of fact, by demanding the equivalence of two factorization theorems (\ref{NRQCD:factorization:prod}) and (\ref{pQCD:factorization:large:pt}), one concludes that the fragmentation function itself
must be subject to the following NRQCD factorization theorem:
%------------------------------------------
\begin{equation}
%------------------------------------------
D_{i\to H}(z) =  \sum_n d^{(n)}(z) \langle {\cal O}^{H}_n \rangle.
%------------------------------------------
\label{NRQCD:factorization:fragmentation:function}
%------------------------------------------
\end{equation}
%------------------------------------------
Here $d^{(n)}(z)$ are the perturbatively calculable coefficient functions, and
$\langle {\cal O}^{H}_n \rangle$ are the same NRQCD matrix elements as
appear in (\ref{NRQCD:factorization:prod}).

In passing, it is worth noting that, for heavy quarkonium production, the NLO power correction (the order-$1/p_T^2$ contribution in  (\ref{pQCD:factorization:large:pt})) also has recently been systematically developed. 
As a consequence, a new set of nonperturbative functions, 
dubbed {\it double-parton fragmentation functions}, must be introduced~\cite{Kang:2011mg,Kang:2014tta}.
Analogous to (\ref{NRQCD:factorization:fragmentation:function}), they are also subject to a similar NRQCD factorization procedure,
with various LO short-distance coefficient functions having been recently calculated~\cite{Ma:2013yla,Ma:2014eja}.

The physical picture underlying (\ref{NRQCD:factorization:fragmentation:function})
was first elucidated and pursued in NRQCD context by Braaten and collaborators in early 90s (Note there were earlier work 
along this direction in the pre-NRQCD era~\cite{Chang:1979nn}).
In those work, various quarkonium fragmentation functions were computed to lowest order in both $\alpha_s$ and $v$, {\it e.g.}
gluon/charm quark fragmentation into $S$-wave quarkonium have been computed~\cite{Braaten:1993rw,Ma:1994zt}.
Recently, the relativistic corrections have been investigated for the $g\to J/\psi$ fragmentation function~\cite{Bodwin:2003wh,Bodwin:2012xc},
as well as for the $c\to J/\psi,\eta_c$ fragmentation functions~\cite{Sang:2009zz}.
Very recently, the order-$\alpha_s$ correction has also
been addressed to the $g\to \eta_c$ fragmentation function~\cite{Artoisenet:2014lpa}.

The aim of the present work is to fill a missing gap, {\it i.e.}, to compute the leading relativistic correction to the $g\to\eta_c$ fragmentation function. This piece of knowledge, in supplement with the recently available radiative correction~\cite{Artoisenet:2014lpa},
might be helpful to interpret the recent LHC measurements on $\eta_c$ production at high $p_T$.

From theoretical perspective, there are many equivalent ways to calculate the quarkonium fragmentation functions.
Originally, Braaten and collaborators have invented a trick to directly extract the fragmentation functions in a process-independent fashion. This method is simple and efficient for a LO calculation, but may become cumbersome if one proceeds to higher order in $\alpha_s$ and $v$. Soon after, Ma pointed out that~\cite{Ma:1994zt}, the NRQCD factorization of quarkonium fragmentation 
function can be conveniently calculated starting
from the operator definition of fragmentation function introduced by Collins and Soper~\cite{Collins:1981uw}.
This elegant approach has the advantage that preserves manifest gauge invariance, and
allows one to systematically address the higher-order corrections, 
as was illustrated in \cite{Bodwin:2003wh,Artoisenet:2014lpa}.

In this work, we will use three different approaches to compute the order-$v^2$ correction to the
$g\to\eta_c$ fragmentation function, {\it i.e.}, Collins-Soper definition, Braaten-Yuan method,
and extracting from a specific physical process involving $\eta_c$ production. 
After obtaining the desired expressions,
we also attempt to study the evolution behavior of the fragmentation functions
of $g\to\eta_c$ and $c\to\eta_c$.

The rest of this paper is organized as follows.
%------------------------------------------
In Section~II, we present a short review on the factorization of the fragmentation function 
for $g\to\eta_c$ in NRQCD framework, and
briefly outline our matching strategy.
%-----------------------------------------
In Section~III, we compute the short-distance coefficients for the gluon fragmenting into the spin-singlet state $\eta_c$ 
through relative order-$v^2$ by employing three different methods, and confirm that all of them yield the identical answer.
%------------------------------------------
In Section~IV, we revisit the $c$ quark fragmenting into $\eta_c$ and verify the previous results through relative order-$v^2$.
We also employ the DGLAP equation to evolve both the fragmentation functions of 
$g\to\eta_c$ and $c\to\eta_c$ to higher energy scales.
%------------------------------------------
Finally we summarize in Section~V.

%------------------------------------------
\section{Fragmentation function in NRQCD factorization and strategy of matching}
%------------------------------------------
\label{FF:NRQCD:fac:strategy:matching}
%------------------------------------------

In line with the NRQCD factorization (\ref{NRQCD:factorization:fragmentation:function}), through the relative order-$v^2$,
the fragmentation function of a gluon fragmenting into the pseudoscalar quarkonium $\eta_c$ reads
%------------------------------------------
\begin{equation}
%------------------------------------------
%\begin{split}
%------------------------------------------
D_{g\to \eta_c} (z)= d^{(0)}(z)\langle \mathcal{O}^{\eta_c}_1 \rangle+ d^{(2)}(z) {\langle \mathcal{P}_1^{\eta_c} \rangle\over m^2} + \mathcal{O}(v^3),
%\end{split}
%------------------------------------------
\label{g:to:etac:NRQCD:factorization}
\end{equation}
%------------------------------------------
where $d^{(0)}(z)$ and $d^{(2)}(z)$ are the corresponding short-distance coefficient functions.
$ \mathcal{O}^{\eta_c}_1 $ and $ \mathcal{P}^{\eta_c}_1 $ are color-singlet NRQCD production operators:
%------------------------------------------
%------------------------------------------
\begin{subequations}
\bqa
%------------------------------------------
%------------------------------------------
&& \mathcal{O}_1^{\eta_c} = \sum_X \chi^\dagger\psi |\eta_c+X \rangle\langle \eta_c+X| \psi^\dagger \chi,
%------------------------------------------
\\
%------------------------------------------
&& \mathcal{P}_1^{\eta_c}=\frac{1}{2}\sum_X \left[\chi^\dagger \psi |\eta_c+X \rangle\langle \eta_c+X| \psi^\dagger \left(-\frac{i}{2} \overleftrightarrow{{\bf D}} \right)^2\chi + \text{h}.\text{c}. \right],
%------------------------------------------
\eqa
%------------------------------------------
\label{Def:NRQCD;prod:operators:eta_c}
\end{subequations}
%------------------------------------------
%------------------------------------------
where $\psi$, $\chi$ are Pauli spinor fields in NRQCD, and $\psi^\dagger \overleftrightarrow{\bf{D}} \chi \equiv \psi^\dagger (\bf{D} \chi) - (\bf{D} \psi)^\dagger \chi$, and $D_\mu$ is the gauge-covariant derivative.

The involved nonperturbative matrix elements in (\ref{g:to:etac:NRQCD:factorization}) are
the vacuum expectation values of those color-singlet NRQCD production operators as specified in (\ref{Def:NRQCD;prod:operators:eta_c}).
Under vacuum saturation approximation, which is accurate up to ${\cal O}(v^4)$, the LO matrix element can be well
approximated by the Schr\"{o}dinger wave functions at the origin for the $\eta_c$ in the potential model:
\begin{equation}
\langle \mathcal{O}_1^{\eta_c}\rangle \approx \frac{N_c}{2\pi} \lvert R_{\eta_c}(0) \rvert^2,
\label{rel:O1:R0square}
\end{equation}
where $N_c=3$ is the number of colors in QCD.

Rather than cope with the order-$v^2$ matrix element itself, it is more convenient to introduce a dimensionless
ratio of the following NRQCD matrix elements:
%------------------------------------------
\beq
%------------------------------------------
\la v^2 \ra_{\eta_c} = \frac{\la \mathcal{P}^{\eta_c}_1\ra}{m^2 \la\mathcal{O}^{\eta_c}_1\ra}\approx\fr{\la \eta_c|\psi^\dag(-\fr{i}{2}\overleftrightarrow{{\bf{D}}})^2\chi|0\ra}{m^2\la \eta_c|\psi^\dag\chi|0\ra},
%------------------------------------------
\eeq
%------------------------------------------
where the second equality is again obtained by invoking the vacuum saturation approximation.
It is often useful to estimate the $\la v^2 \ra_{\eta_c}$ from the so-called Gremm-Kapustin relation~\cite{Gremm:1997dq}:
%------------------------------------------
\beq
%------------------------------------------
\fr{M_{\eta_c}}{2m}=1+\fr{1}{2} \la v^2 \ra _{\eta_c}+O(v^4).
%------------------------------------------
\label{gkrltn}
\eeq
%------------------------------------------
If the charm quark mass is taken as the one-loop pole mass, $m=1.4$ GeV, then $\la v^2 \ra _{\eta_c}\approx 0.13$.

Our central goal is to determine the short-distance coefficient functions $d^{(0)}(z)$ and $d^{(2)}(z)$.
To this purpose, we will use the standard matching technique. Namely, since the short-distance coefficients are independent of the long-distance dynamics, we can freely replace the physical hadron $\eta_c$ by a free
$c(p)\bar{c}(\bar{p})$ pair carrying the following momenta:
%------------------------------------------
\begin{equation}
%------------------------------------------
p= \dfrac{P}{2} + q,  \qquad \bar{p} =\dfrac{P}{2}  - q,
%------------------------------------------
\end{equation}
%------------------------------------------
where $P^2=4 E^2$, $P\cdot q=0$ and $q^2= m^2-E^2$, so that $p^2={\bar p}^2=m^2$.
In the $c\bar{c}$ pair (quarknoum) rest frame, one has $P^\mu=(2E,\bf{0})$, $q^\mu=(0,\bf{q})$,
so $E=\sqrt{m^2+\bf{q}^2}$. 
We can further enforce the $c\bar c$ pair to bear quantum number ${}^1S_0^{(1)}$.

We then substitute this fictitious $\eta_c$ state into the factorization formula (\ref{g:to:etac:NRQCD:factorization}):
%------------------------------------------
\begin{equation}
%------------------------------------------
D_{g\to c\bar c({}^1S_0^{(1)})} (z)= d^{(0)} (z)  \la \mathcal{O}_1^{c\bar c({}^1S_0^{(1)})}
\ra+ d^{(2)}(z) {\la \mathcal{P}_1^{c\bar c({}^1S_0^{(1)})}\ra \over m^2} +  \mathcal{O}(v^3),
%------------------------------------------
\label{g:to:free:ccbar:pair:NRQCD:factorization}
\end{equation}
%------------------------------------------
Since now both the left-hand side and right-hand side in (\ref{g:to:free:ccbar:pair:NRQCD:factorization})
can be computed in perturbation theory, we can readily solve for $d^{(0)}(z)$ and $d^{(2)}(z)$.

Firstly, one can trivially deduce the NRQCD matrix elements that appear in (\ref{g:to:free:ccbar:pair:NRQCD:factorization}):
%------------------------------------------
\begin{subequations}
%------------------------------------------
\bqa
%------------------------------------------
&& \la \mathcal{O}_1^{c\bar c({}^1S_0^{(1)})}\ra =2N_c,\\
%------------------------------------------
&& \la \mathcal{P}_1^{c\bar c({}^1S_0^{(1)})}\ra = 
\la \mathcal{O}_1^{c\bar c({}^1S_0^{(1)})} \ra {\bf q}^2.
%------------------------------------------
%------------------------------------------
\eqa
%------------------------------------------
\end{subequations}
%------------------------------------------
Note that the $c(\bar{c})$ state in the NRQCD matrix elements obeys the nonrelativistic normalization.

We briefly describe our strategy of computing the left-hand side of (\ref{g:to:free:ccbar:pair:NRQCD:factorization}).
After writing down the QCD amplitude to produce the free $c(p)$ and $\bar{c}(\bar{p})$,
$\bar{u}(p) {\cal A}v(\bar{p})$,
we have to project out the $c(p)\bar{c}(\bar{p})$ pair onto the desired ${}^1S_0^{(1)}$ state.
We employ the standard covariant trace technique~\cite{Bodwin:2002hg}, with the aid of  
the following projector:
%------------------------------------------
\begin{equation}
 \Pi_1^{(1)}= \dfrac{( \bar{p}\!\!\!/ -m ) \gamma_5 (P\!\!\!\!/ + 2 E) ( p\!\!\!/ + m)}{8\sqrt{2}E^2(E+m)}\otimes
%------------------------------------------
\dfrac{1_C}{\sqrt{N_c}}.
\label{color:spin:projector}
%------------------------------------------
\end{equation}
%------------------------------------------
Thereby we extract the singlet amplitude by the operation ${\cal M}= \bar{u}(p) {\cal A}v(\bar{p})
 \to {\rm Tr({\cal A} \Pi_1^{(1)})}$, where the $c\bar{c}$ pair is in the spin/color-singlet state.
We emphasize that, by using the above projection operator in \label{color:spin:projector}, one has tacitly assumed that
the quark and antiquark in the QCD side are normalized nonrelativistically.

We need to further single out the $S$-wave orbital angular momentum contribution. We first
truncate the amplitude $\cal M$ to quadratic order in $q^\mu$, then take the following procedure:
%------------------------------------------
\beq
%------------------------------------------
\mathcal{M}_{S{\rm \!-wave}} = \mathcal{M}_0 +
\dfrac{\bold{q}^2}{m^2} \mathcal{M}_2 + \mathcal{O} ({\bf{q}}^4),
%------------------------------------------
\label{eq:qnexp1}
\eeq
%------------------------------------------
where
%-----------------------------------------
\begin{subequations}
%-----------------------------------------
\bqa
%------------------------------------------
 &&\mathcal{M}_0=\lim\limits_{q\to 0} \mathcal{M},
%------------------------------------------
 \\
%------------------------------------------
&& \mathcal{M}_2 =\dfrac{m^2}{6}\left(-g^{\alpha \beta} +\dfrac{P^\alpha P^\beta}{4E^2} \right) \lim\limits_{q\to 0} \left( \dfrac{\partial^2 \mathcal{M}}{\partial q^\alpha \partial q^\beta} \right).
%------------------------------------------
\eqa
%------------------------------------------
\label{eq:qnexp2}
\end{subequations}
%------------------------------------------
The $\mathcal{M}_0$ and $\mathcal{M}_2$ are then the desired QCD amplitudes to produce a $c\bar{c}$ pair in the ${}^1S_0^{(1)}$ state, accurate through order-$v^2$.

%%%%%%%%%%%%%%%%%%%%%%%%%%%%%%%%%%%%%%%%%%%%%%%%%%%%%%%%%%%%%%%%%%%%%%%%%%%%%%
\section{Perturbative calculation of short-distance coefficients through order-$v^2$}
%%%%%%%%%%%%%%%%%%%%%%%%%%%%%%%%%%%%%%%%%%%%%%%%%%%%%%%%%%%%%%%%%%%%%%%%%%%%%%
\label{Perturbative:calc:short-distance}
%%%%%%%%%%%%%%%%%%%%%%%%%%%%%%%%%%%%%%%%%%%%%%%%%%%%%%%%%%%%%%%%%%%%%%%%%%%%%%

In this section, we will employ several different methods to ascertain the short-distance coefficients associated with the gluon fragmenting into $\eta_c$ through order $v^2$.
They all yield identical results, thus serving as a useful consistence check.

\subsection{From Collins-Soper definition}

The rigorous operator definition of the fragmentation function
was introduced by Collins and Soper in 1981~\cite{Collins:1981uw}.
It is convenient to adopt the light-cone coordinate, where $x^\mu= (x^+, x^-, \bold{x}_\bot)$
with $x^\pm=(x^0\pm x^3)/\sqrt{2}$, and $\bold{x}_\bot = (0,x^1, x^2,0)$.
We assume the parent (virtual) gluon moves along the $z$ axis, that is, $k^\mu =(k^+, k^-, \bold{k}_\bot={\bold 0})$,
and the final-state hadron $H$ carries the momentum $P^{\mu}=(P^+, P^-={M_H^2+{\bf P}_\perp^2 \over 2 P^+},
{\bold P}_\bot)$, where $M_H$ is the hadron mass.
The corresponding $g\to H$ fragmentation function is defined as
%------------------------------------------

%------------------------------------------
\begin{align}
%------------------------------------------
D_{g\to H}(z)=&\frac{- g_{\mu\nu} z^{d-3}} {2\pi k^+ (N^2_c-1) (d-2)}\!\!\int_{-\infty}^{+\infty}\!\!\! dx^-
e^{-ik^+ x^-}\!\!\sum_X  \langle 0| G^{+\mu}_{a}(0) \mathcal{L}\left[0,\infty\right]_{ab}
|H(P^+\!,\!P^-\!,\! {\bf P}_\perp)\!+\!X\rangle
%------------------------------------------
\nn\\
%------------------------------------------
&\times
\langle H(P^+,P^-, {\bf P}_\perp)+X|\mathcal{L}\left[\infty,x^-\right]_{bc}
 G^{+\nu}_{c}(x^-) |0\rangle,
\label{CSdef}
\end{align}
where $d=4$ is the spacetime dimension, and $G^{+\mu}_a$ signifies the gluon field strength tensor.
Specifically, the daughter hadron $H$ carries the 4-momentum
$(P^+=z k^+, P^-= {M_H^2+{\bf P}^2_\bot \over 2 z k^+},{\bf P}_\perp)$.
$\mathcal{L}\left[x^-,y^-\right]_{ab}$
represents the gauge link:
%------------------------------------------
%------------------------------------------
\bqa
%------------------------------------------
&& \mathcal{L}\left[x^-,y^-\right]_{ab}=\left[ \mathcal{P} \exp \left(\!\! -i g_s \int_{x^-}^{y^-} d\xi^-  n\cdot A(0,\xi^-,\bold{0}_\bot) \!\! \right) \right]_{ab},
%------------------------------------------
\eqa
%------------------------------------------
%------------------------------------------
where $A^\mu=A^\mu_a T^a$ is the matrix-valued gluon field, and $T^a$ is the generator of the $SU(N_c)$ group in adjoint representation. $\mathcal{P}$ implies the path-ordering.
The null vector $n^\mu=\left(0,1,\bf 0_\bot\right)$ defines the ``minus"
light-cone direction, so that $n\cdot A=A^+$.

%---------------------------
\begin{figure}[H]
\begin{center}
\includegraphics[scale=0.35]{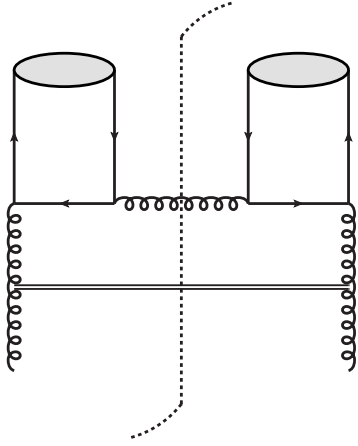}
 \caption{One typical diagram for the fragmentation function $g\to c\bar{c}({}^1S_0^{(1)})$ at the leading order
 in $\alpha_s$ in Feynman gauge.
 The double line represents the eikonal line that originates from the gauge link, the vertical dashed line
 implies imposing a cut.
\label{FDCoSo} }
\end{center}
\end{figure}
%--------------------------

Following the above definition, together with the matching strategy outlined in Sec.~2,
we can replace the physical $\eta_c$
by a fictitious free $c\bar{c}$ pair.
For the gluon fragmenting into the $c\bar{c}({}^1S_0^{(1)})$ state,
in Feynman gauge one can draw four Feynman diagrams at the leading order in $\alpha_s$,
one of which is depicted in Fig.~{\ref{FDCoSo}}.
The relevant Feynman rules can be found in Ref.~\cite{Collins:1981uw}.
After some algebra, this perturbatively calculable 
fragmentation function can be expressed as
%--------------------------
\begin{align}
%--------------------------
\notag D_{g\to c\bar{c}{(^1S_0)}}=& \frac{{\alpha_s^2 E}}{N_c^2-1}\int\frac{dl^+d^2 \bold{l}_{\bot}}{(2\pi)^3 2 l^+}\int\frac{dP^+d^2 \bold{P}_\bot}{(2\pi)^3 2 P^+}2\pi\delta(zk^+-P^+)\delta(k^+ - P^+-l^{+})\\
&\times\delta^{(2)}(\bold{P}_\perp+\bold{l}_\perp) F_c\sum^2_{i, j =1}\text{Tr}\left[ L^{\sigma\alpha}_i \mathcal{G}_{\rho \sigma \alpha\beta}(L^{\rho\beta}_j)^\dagger\right] ,
%--------------------------
\label{frag:function:Collins:Soper}
\end{align}
%--------------------------
where $F_c = \text{Tr} (T^aT^b)\text{Tr}(T^aT^b) ={N_c^2-1\over 4}$ is the corresponding color factor, 
$l^\mu$ stands for the 4-momentum of the gluon recoiling against the $c\bar{c}$ pair.
The rank-4 tensor $\mathcal{G}_{\rho \sigma \alpha\beta}=g_{\alpha\beta} g^{\mu\nu}(g_{\mu\rho}k^+ - k_\mu n_\rho )
( g_{\nu\sigma}k^+ - k_\nu n_\sigma )$ stems from the product of two vertices
of gluon interacting with the eikonal line, while $L_{i}$ ($i=1,2$) representing the remaining ordinary 
quark-gluon amplitudes:
%--------------------------
\begin{subequations}
%--------------------------
\begin{align}
%--------------------------
\label{lfunctions}
%--------------------------
&L_1^{\sigma\alpha} \!= \! \frac{\text{Tr} \left[ \gamma^\sigma (\frac{1}{2} P\!\!\!\!/ + q\!\!\!/ +l\!\!\!/ + m ) \gamma^\alpha \Pi_1^{(1)} \right]}{(\frac{P}{2}+q+l )^2 - m^2} ,
%--------------------------
\\
%--------------------------
&
L_2^{\rho\beta} \!=\! \frac{\text{Tr} \left[ \gamma^\beta ( - \frac{1}{2} P\!\!\!\!/ + q\!\!\!/ - l\!\!\!/ + m )
\gamma^\rho \Pi_1^{(1)}  \right]}{(\frac{P}{2} - q+l )^2 - m^2},
%--------------------------
\end{align}
\end{subequations}
%--------------------------
%--------------------------
where the projector $\Pi_1^{(1)}$ has been given in (\ref{color:spin:projector}).

The $S$-wave amplitudes are extracted following the recipe given in (\ref{eq:qnexp1}) and (\ref{eq:qnexp2}).
After squaring the $S$-wave amplitudes, carrying out the phase-space integration to get rid of all the $\delta$-functions in (\ref{frag:function:Collins:Soper}),
and finally performing the trivial angular
integration in $\bf{P}_\perp$, we end up with the following one-fold integral:
%--------------------------
%
\begin{equation}
D_{g\!\to\! c\bar{c}{(^1S_0^{(1)})}}\!\!=\! {\alpha_s^2 } \!\!\int_0^\infty d \rho \left( \mathcal{A}_0+  {{\bf q}^2\over m^2} \mathcal{A}_2 +{\cal O}({\bf q}^4)\right).
\label{g:to:free:ccbar:pair:NRQCD:factorization:cs}
\end{equation}
%--------------------------
%
%--------------------------
where $\rho = \bold{P}_\perp^2$ is the modulus of the transverse momentum of the $c\bar{c}$ pair, and
%--------------------------

%
\begin{equation}
\label{analytical:A0}
%--------------------------
\mathcal{A}_0 = \frac{8 z (1-z)  \Big(\rho^2\left(z^2+(1-z)^2\right) + 8 \rho m^2 (1-z)^3 +16 m^4(1-z)^4\Big)}{m \text{$N_c$} \left(4 m^2 (1-z)+\rho \right)^2 \left(4 m^2 (1-z)^2+\rho \right)^2},
\end{equation}
%--------------------------
%\
%--------------------------
%--------------------------
and
%--------------------------
%--------------------------
\begin{align}
\notag \mathcal{A}_2\! =\! & -\frac{ 4 z (1\!-\!z)} {3 N_c m \left(4 m^2 (1\!-\!z)\!+\!\rho \right)^3\! \left(4 m^2 (1\!-\!z)^2+\rho \right)^3}\Big[5 \rho^4\left(z^2\!+\!(1\!-\!z)^2\right)\\
\notag &-\! 4 \rho^3 m^2 (1\!-\!z)(34z^3\! - \! 100z^2\!+\!81z \!-\! 32)+16 \rho^2 m^4(1\!-\! z)^3(80z^2\!-\! 117z\!+\! 66) \\
& -64\rho m^6(1\!-\!z )^5(51z\! -\! 56)+4352 m^8 (1\!-\! z)^7 \Big].
\label{analytical:A2}
\end{align}

The integral over $\rho$ can be transformed into more conventional phase-space integral through
%--------------------------
%--------------------------
\beq
%--------------------------
\rho = (1-z) (z s -4 E^2),
%--------------------------
\eeq
%--------------------------
%--------------------------
where $s$ is the squared invariant mass of the final-state $c\bar c +g$ system.
$\rho\ge 0$ then implies that $s\geq 4E^2/z$. If we replace $\rho$ by $s$ in (\ref{g:to:free:ccbar:pair:NRQCD:factorization:cs}),
and replace the lower boundary to $s\geq 4m^2/z$, we can fully recover the corresponding LO
expression in \cite{Artoisenet:2014lpa}.

After fulfilling the integration over $\rho$, and matching (\ref{g:to:free:ccbar:pair:NRQCD:factorization}) onto (\ref{g:to:free:ccbar:pair:NRQCD:factorization:cs}),
we can obtain the corresponding short-distance coefficient functions accurate
through order-$v^2$:
%--------------------------
%--------------------------
\begin{subequations}
%--------------------------
\bqa
%--------------------------
&& d^{(0)}(z) = \frac{\alpha_s^2 }{4N^2_c m^3}  \left[ 3z -2z^2+ 2(1-z) \text{ln}(1-z) \right],
\label{d0:analy:expression}
%--------------------------
\\
%--------------------------
&& d^{(2)}(z) = -{11 \over 6} d^{(0)}(z).
\label{d2:analy:expression}
%--------------------------
\eqa
%--------------------------
\label{d0:d2:analytical:expressions}
\end{subequations}
%--------------------------
%-------------------------- 
Eq.~(\ref{d0:analy:expression}) recovers the well-known LO result~\cite{Braaten:1993rw,Ma:1994zt}.
Eq.~(\ref{d2:analy:expression}) is the central result of this work. 
The relativistic correction tends to dilute the LO fragmentation contribution.
A curious feature is that,
despite the integrand in (\ref{analytical:A2}) does not resemble (\ref{analytical:A0}) at all,
$d^{(2)}(z)$ turns out to bear the exactly identical functional dependence on $z$
as $d^{(0)}(z)$. 
It is interesting to observe that, likely to be a sole coincidence,
for the short-distance coefficient function associated with $g\to c\bar{c}({}^{3}S_1^{(8)})$,
the ratio of the order-$v^2$ term and the order-$v^0$ term also turns out to
be $-{11\over 6}$~\cite{Bodwin:2003wh}.

We have also redone the calculation in the light-cone gauge $A^+=0$, where the gauge link in Fig.~\ref{FDCoSo} is absent.
We again reproduce the results listed in (\ref{d0:d2:analytical:expressions}).
Hence, we have explicitly checked the gauge invariance of this fragmentation function according to the
Collins-Soper definition.

Substituting these short-distance coefficients (\ref{d0:d2:analytical:expressions}) into the NRQCD factorization formula (\ref{g:to:etac:NRQCD:factorization}), we then obtain the fragmentation function of $g\to \eta_c$.
It is interesting to deduce the total fragmentation probability of $g\to \eta_c$ (the 1st Mellin moment of the fragmentation function):
%--------------------------
\beq
%--------------------------
\int^1_0 dz\,D_{g\to \eta_c} (z)
= {\alpha _s^2  \langle {\mathcal O}^{\eta_c}_1 \rangle \over 12 N_c^2 m^3}
\left(1-{11\over 6} \langle v^2\rangle_{\eta_c} \right),
%--------------------------
\label{eq:g_FrgmnttnPrblty}
\eeq
%--------------------------
Obviously, provided that $\langle v^2\rangle_{\eta_c}$ is positive,
incorporating the relativistic correction decreases 
the fragmentation probability of $g\to \eta_c$.

For the latter use, we are also interested in the 2nd Mellin moment of the fragmentation function,
which might be interpreted as the average momentum fraction of the $\eta_c$ meson in $g$ fragmentation process:
%--------------------------
\beq
%--------------------------
\int^1_0 dz\,z D_{g\to \eta_c} (z)
= {\alpha _s^2  \langle {\mathcal O}^{\eta_c}_1 \rangle \over 18 N_c^2 m^3}
\left(1-{11\over 6} \langle v^2\rangle_{\eta_c} \right).
%--------------------------
\label{eq:g_Frgmnttn_2nd_mmnt}
\eeq
%--------------------------

\subsection{Extraction from Higgs boson decay}

In this section, we attempt to extract the $g\to \eta_c$ fragmentation function from a specific
process, say, inclusive $\eta_c$ production from Higgs boson decay, $h\to g^* g \to \eta_c+gg$.

The Higgs coupling to two gluons plays a crucial role in discovering the Higgs boson at the LHC experiment.
In Standard Model, it can be represented by an effective operator $-{\lambda\over \texttt{v}} h G^a_{\mu \nu}G^{a,\mu \nu}$,
where $\texttt{v}$ signifies the Higgs vacuum expectation value, and the effective coupling 
$\lambda={\alpha_s\over 12\pi}+{\cal O}(\alpha_s^2)$ receives the major contribution from the top quark loop.
%The corresponding Feynman rule for this effective operator reads ${-\fr{2\lambda}{v}(g^{\mu \nu}k_1 \cdot k_2- k_1^\mu %k_2^\nu)}\delta^{ab}$. From this Feynman rule,
From this effective operator, one can readily deduce the Higgs boson hadronic width
$\Gamma_0= \frac{2\lambda^2 M_h^3}{ \pi \texttt{v}^2}$.

We are interested in inferring the energy spectrum of the $\eta_c$ meson in $h(K)\to \eta_c(P)+g(k_1)g(k_2)$.
It is convenient to introduce the three dimensionless energy fraction variables:
%--------------------------
\begin{subequations}
%--------------------------
\bqa
%--------------------------
&& z \equiv {2P\cdot K\over K^2} = {2P^0\over M_h},\\
%--------------------------
&& x_1 \equiv {2k_1\cdot K\over K^2} = {2k_1^0\over M_h},\\
%--------------------------
&& x_2 \equiv {2k_2\cdot K\over K^2} = {2k_2^0\over M_h},
%--------------------------
\eqa
\end{subequations}
%--------------------------
which is subject to the energy conservation condition $x_1+x_2+z=2$. 
For convenience, we also introduce a dimensionless ratio $r\equiv {M_{\eta_c^2}\over M_h^2}$. 
We then expect that the gluon fragmentation function can be read off from
%--------------------------
%--------------------------
\begin{equation}
%--------------------------
  D_{g\rr \eta_c}(z) = \lim_{M_h\to \infty} {1\over\Gamma_0}\,\fr{d\Ga [h \rr \eta_c(z)+gg]}{dz},
%--------------------------
\label{frag:func:higgs:decay}
%--------------------------
 \end{equation}
%--------------------------
while holding $M_{\eta_c}$ fixed.

\begin{figure}[H]
  \centering
  \includegraphics[width=0.3\linewidth]{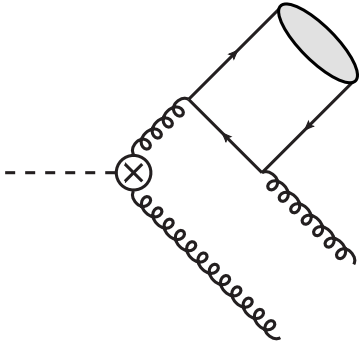}
 \protect\caption{One typical diagram for $h \to c\bar{c}(P,{}^1S_0^{(1)})+gg$.
The cross denotes the $h G^a_{\mu \nu}G^{a,\mu \nu}$ vertex.
We suppress the additional three diagrams, which can be obtained by reversing the quark line 
and permutating the final state gluons. }\label{hggetac}
\end{figure}

In accordance with the matching ansatz, our goal is  again to compute the process $h \to c\bar{c}(P,{}^1S_0^{(1)})+gg$.
At the lowest order in $\alpha_s$, there are four Feynman diagrams, one of which is depicted in Fig.~\ref{hggetac}.

After truncating the $S$-wave amplitude through order ${\bf q}^2$, we 
then obtain the squared amplitude according to
$\lvert \mathcal{M} \rvert^2 = |\mathcal{M}_0|^2 +
{{\bf q}^2\over m^2} (\mathcal{M}_0 \mathcal{M}_2^*+ \mathcal{M}_2 \mathcal{M}_0^* )$. 
Accordingly, the decay rate of $h \to c\bar{c}({}^1S_0^{(1)})+gg$ can be put in the form:
%--------------------------
\begin{equation}
%--------------------------
\Gamma[h\to c\bar{c}(z,{}^1S_0^{(1)})+ gg] = \frac{M_h }{256\pi^3} \int^{1+r}_{2\sqrt{r}} dz \int^{x_1^+}_{x_1^-} dx_1 \lvert \mathcal{M} \rvert^2,
%--------------------------
\end{equation}
%--------------------------
where we have replaced $r$ by $4E^2/M_h^2$. The integration boundaries of $z$ are explicitly labeled.
Obviously, its allowed range reduces to
$0\le z\le 1$ as $M_h \to \infty$.
The boundaries for the energy fraction of gluon 1, dubbed $x_1^\pm$, read
%--------------------------
\begin{equation}
%--------------------------
x_1^{\pm}= {2-z\pm \sqrt{z^2-4r} \over 2}.
%--------------------------
\label{x1:upper:lower:bound}
%--------------------------
\end{equation}
%--------------------------

After carrying out the phase-space integration over $x_1$ and substituting (\ref{x1:upper:lower:bound}), we can reshuffle
the corresponding expression to the second order in ${\bf q}$.
In line with (\ref{frag:func:higgs:decay}), dividing this expression by the hadronic decay width of $h\to gg$, and
we only retain those terms that survive in the $r\to 0$ limit.
By solving the matching equation (\ref{g:to:free:ccbar:pair:NRQCD:factorization}),
we find the exactly identical expressions of
$d^{(0)}(z)$ and $d^{(2)}(z)$ as given in (\ref{d0:d2:analytical:expressions}),
which were previous obtained via Collins-Soper definition.

Alternatively, one can also apply the Braaten-Yuan trick~\cite{Braaten:1993rw} to extract
the gluon fragmentation function, through order $v^2$. 
In similar spirit to the preceding Higgs boson decay example, 
this approach also aims to extract the fragmentation function from a physical process. Nevertheless, 
the virtue of this method is that it does not need specify any concrete process, 
and the only required knowledge is the off-shell amplitude $g^* \to \eta_c + g$. 
After some trick in factorizing the phase space integration, one ends up with
a one-dimensional integral which exactly resembles what is encountered in the Collins-Soper approach.
Not surprisingly, we again reproduce the results for $d^{0}(z)$ and $d^{2}(z)$ as given 
in (\ref{d0:d2:analytical:expressions}).

\section{Evolution of fragmentation functions}
\label{Evolution:frag:functions}

Recently LHC has measured the $\eta_c$ production around $p_T\le 16\  \text{GeV}$~\cite{Aaij:2014bga}, which is already
considerably greater than the charm quark mass. One may worry that the large collinear logarithm
$(\alpha_s \ln{p_T^2 \over m_c^2})^n$ could potentially ruin the fixed-order calculation.
These large logarithms are most conveniently resummed by invoking the famous DGLAP equation:
%--------------------------
%--------------------------
\begin{equation}
%--------------------------
\mu \frac{\partial}{\partial \mu} D_{i\to \eta_c} (z, \mu) =\fr{\alpha_s(\mu^2)}{\pi} \sum_{j\in(g,c)} \int^1_z \frac{dy}{y} P_{j\leftarrow i}\left(\frac{z}{y}, \mu\right) D_{j\to \eta_c}(y, \mu),
%--------------------------
\end{equation}
%--------------------------
%--------------------------
where $P_{j\leftarrow i}$ is the splitting kernel for parton $i$ splitting into parton $j$.
Therefore, in order to understand the evolution of $g\to\eta_c$ fragmentation function, we inevitable also need the knowledge
on the $c\to\eta_c$ fragmentation function, due to the nonvanishing off-diagonal splitting kernel $P_{j\leftarrow i}$ for $i \neq j$.
For simplicity, we neglect the contribution from the light quark/antiquark fragmenting
into $\eta_c$, since they are suppressed by additional powers in $\alpha_s$.

The LO splitting kernels read:
\begin{subequations}
\begin{align}
P_{c\leftarrow c}(z)&=\fr{4}{3}\left[\fr{1+z^2}{\left(1-z\right)_+}+\fr{3}{2}\delta\left(1-z\right) \right] ,\\
P_{g\leftarrow c}(z)&=\fr{4}{3}\left[\fr{1+\left(1-z\right)^2}{z} \right],\\
P_{c\leftarrow g}(z)&=\fr{1}{2}\left[z^2+\left(1-z\right)^2  \right],\\
P_{g\leftarrow g}(z)&=6\Big{[}  \fr{\left(1-z\right)}{z}+\fr{z}{\left(1-z\right)_+}+z\left(1-z\right)+\left(\fr{11}{12}-
\fr{n_f}{18}\right)\delta\left(1-z\right)\Big{],}
\end{align}
\end{subequations}
where $n_f$ is the number of active light quark flavors.

The LO fragmentation function of the $c$ quark into $\eta_c$ was known long ago~\cite{Braaten:1993mp,Ma:1994zt}.
The first-order relativistic correction has been computed recently~\cite{Sang:2009zz}. 
Through order $v^2$, the corresponding short-distance
coefficient functions read
%___________________________________
\begin{subequations}
\begin{align}
&d^{(0)}_{c\rr\eta_c}(z) =\frac{16 \alpha_s^2 z(1-z)^2(48+8z^2- 8z^3+3z^4)}{243 m^3(2-z)^6},  \\
&d^{(2)}_{{c\rr\eta_c}}(z) = \frac{8 \alpha_s^2 z(\!1\!-\!z\!)^2\!}{729 m^3(2-z)^8}(-2112\!+\!\!2496z\!-\!80z^2\!\!+\!\!128z^3\!-\!268z^4\!\!+\!\!148z^5\!-\!15z^6).
\end{align}
\label{d0:d2:c:frag:etac:analyt}
\end{subequations}
%-----------------------------------------------
We have revisited this fragmentation function through order-$v^2$ by utilizing two different approaches:
starting from Collins-Soper definition, as well as extracting from a specific process $\gamma^* \to \eta_c + c\bar{c}$.
We have confirmed both the order-$v^0$ and order-$v^2$ results given in (\ref{d0:d2:c:frag:etac:analyt}).

The fragmentation probability for $c \to \eta_c$ is
\beq
%--------------------------
\int^1_0 dz\,D_{c\to \eta_c} (z)
= {8\alpha _s^2  \langle {\mathcal O}^{\eta_c}_1 \rangle \over 81 N_c^2 m^3}
\left({2319\over 5}-666\ln 2 -\left({{1027\over 14}-109\ln 2}\right) \langle v^2\rangle_{\eta_c} \right).
%--------------------------
\label{eq:c_FrgmnttnPrblty}
\eeq
%--------------------------
Here the relativistic correction also tends to dilute the fragmentation probability.

We also compute the 2nd Mellin moment for the $c\to \eta_c$ fragmentation function:
\beq
%--------------------------
\int^1_0 dz\,z D_{c\to \eta_c} (z)
= {8\alpha _s^2  \langle {\mathcal O}^{\eta_c}_1 \rangle \over 81 N_c^2 m^3}
\left({9738\over 5}-2808\ln 2 -\left({{67407\over 35}-2780\ln 2}\right) \langle v^2\rangle_{\eta_c} \right).
%--------------------------
\label{eq:c_Frgmnttn_2nd_mmnt}
\eeq

For the nonperturbative input parameters, we take $\langle \mathcal{O}_1^{\eta_c}\rangle\approx0.244$ $\rm{GeV}^3$,
which is obtained from $\lvert R_{\eta_c}(0) \rvert^2=0.512\;\rm{GeV}^3$~\cite{Braaten:1993mp} through  (\ref{rel:O1:R0square}).
We also take $\langle v^2 \rangle_{\eta_c} = 0.13$, which is obtained through Gremm-Kapustin relation by choosing the one-loop
charm quark pole mass to be 1.4 GeV. The QCD coupling constants at initial scales are set as $\alpha_s(2m_c)=0.266$, $\alpha_s(3m_c)=0.233$ associated with the gluon and the $c$ quark fragmentation, respectively.

We utilize the elaborate FORTRAN/C++ package \textsf{HOPPET}~\cite{Salam:2008qg} to numerically solve the DGLAP evolution equation.
We consider two types of evolution equations: the evolution by only implementing the diagonal splitting kernels ($c\to c$ and $g \to g$) as the non-singlet (NS) (without mixing), as well as the evolution incorporating off-diagonal splitting kernel as singlet (S) (with mixing effect). The starting scales for the non-singlet evolution are set as $3m_c$ and $2m_c$, 
for $c^*\to\eta_c$ and $g^*\to\eta_c$ fragmentation functions, respectively, 
corresponding to the invariant masses of the final sate of $c\to \eta_c+c$ and $g\to \eta_c+g$.
We choose the evolved scales to be $15\ \text{GeV}$ and $45\ \text{GeV}$, respectively.
For the singlet evolution, the starting scale is chosen as $3m_c$, so we take the strategy by carrying out non-singlet evolution of the gluon fragmentation from $2m_c$ to $3m_c$ in the beginning, then perform the 
singlet evolution from $3m_c$ to the scales $15\ \text{GeV}$ and $45\ \text{GeV}$.

%--------------------------
\begin{figure}[H]
\begin{center}
\includegraphics[width=\linewidth]{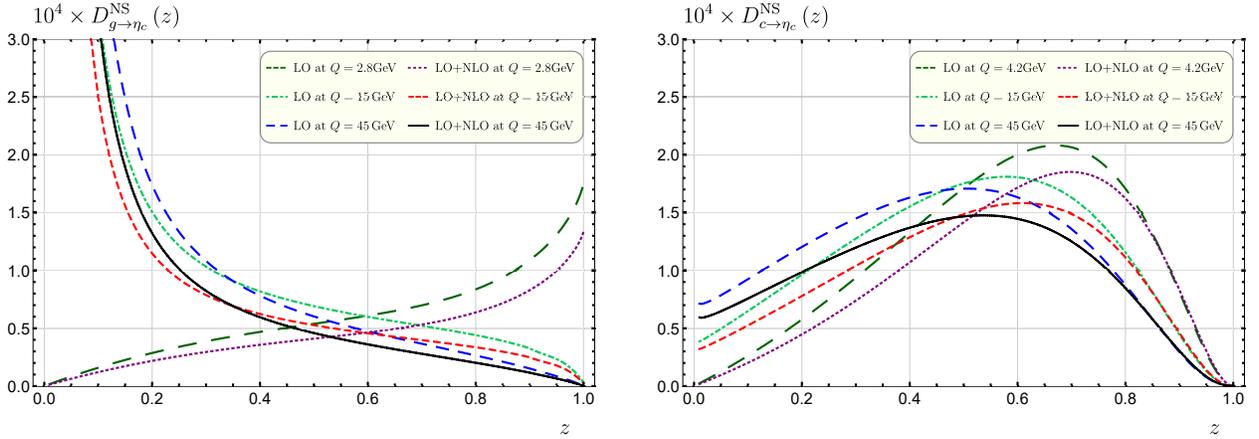}
 \caption{Fragmentation functions of $g\to \eta_c$ and $c\to \eta_c$
 at different energy scale (without mixing).
\label{EvNoM} }
\end{center}
\end{figure}
%--------------------------
%--------------------------
\begin{figure}[H]
\begin{center}
\includegraphics[width=1\linewidth]{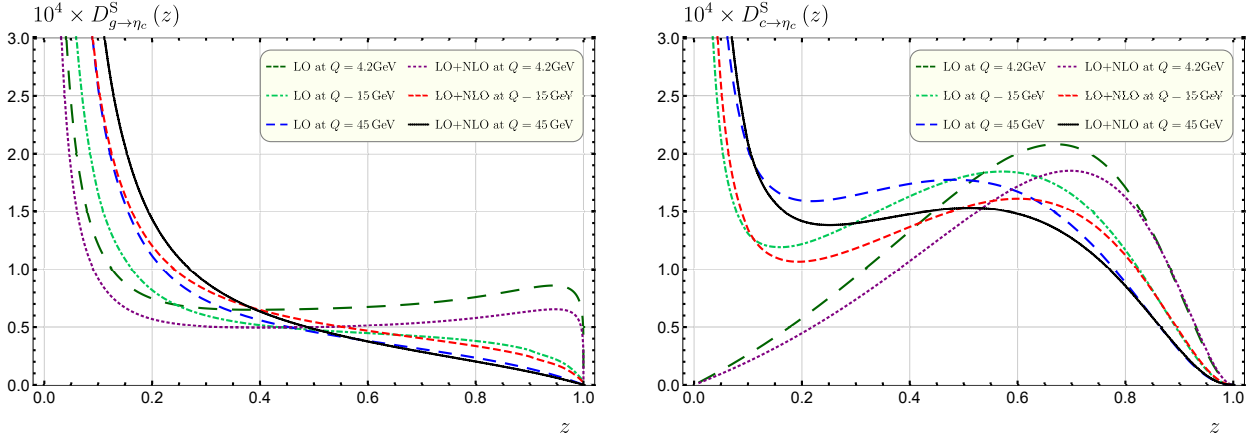}
 \caption{Fragmentation functions of $g\to \eta_c$ and $c\to \eta_c$
 at different energy scale (with mixing).
\label{EvM} }
\end{center}
\end{figure}
%--------------------------

In Fig.~\ref{EvNoM} and Fig.~\ref{EvM}, we display the evolution of the fragmentation functions
with different energy scales, also including relativistic correction effects.
The effect of relativistic corrections at different scale is estimated by the following factor:
%\begin{footnotesize}
\begin{align}
\Delta_{c/g\to \eta_c}\left(Q\right)=\frac{\int_0^1 dz\,z\left[D_{c/g\to \eta_c}\left(z,Q\right)-D^{(0)}_{c/g\to \eta_c}\left(z,Q\right)\right]}
{\int_0^1 dz\,z D^{(0)}_{c/g\to \eta_c}\left(z,Q\right)}.
\end{align}
%\end{footnotesize}
where $D^{(0)}$ is the LO fragmentation function, and $D$ is the LO+NLO fragmentation function. 
The reason why we choose the second Mellin moments instead of the first Mellin moments (fragmentation probability, the results at initial scale is given in Eq.(\ref{eq:g_FrgmnttnPrblty}, \ref{eq:c_FrgmnttnPrblty})) is that the fragmentation functions diverge  quickly in small $z$ region, although first Mellin moments are finite but they largely depend on the contribution from the small $z$ region, where the numerical extrapolation of the evolution is not reliable. 
Therefore, the numerical results of the first Mellin moment are not stable as $z$ gets closer to $0$. 
For the second Mellin moments, the contribution from small $z$ region is suppressed by the $z$ factor and the numerical results are stable.

The numerical results of $\Delta$ are listed in Table \ref{Tb:rltv_crctn}. Form the table we can tell that the $\mathcal{O}\left(v^2\right)$ corrections are around $8\%\sim24\%$ with respect to the $\mathcal{O}\left(v^0\right)$ results. 
It should be noticed that $\Delta_{g\rightarrow\eta_{c}}^{{\rm NS}}\left(Q\right)$ is independent of evolution scale, 
this is because the NLO correction of $g\to \eta_c$ fragmentation function is proportional to the LO result, the ratio is $-\tfrac{11}{6}\langle v^2 \rangle_{\eta_c}=-0.238$.
\begin{table}[H]
%\begin{footnotesize}
\begin{centering}
\begin{tabular}{|c|c|c|c|}
\hline
 & $Q=4.2\rm{GeV}$ & $Q=15{\rm GeV}$ & $Q=45{\rm GeV}$\tabularnewline
\hline
\hline
$\Delta_{c\rightarrow\eta_{c}}^{{\rm NS}}\left(Q\right)$ & $-0.108$ & $-0.108$ & $-0.108$\tabularnewline
\hline
$\Delta_{g\rightarrow\eta_{c}}^{{\rm NS}}\left(Q\right)$ & $-0.238$ & $-0.238$ & $-0.238$\tabularnewline
\hline
$\Delta_{c\rightarrow\eta_{c}}^{{\rm S}}\left(Q\right)$  & $-0.108$ & $-0.093$ & $-0.080$\tabularnewline
\hline
$\Delta_{g\rightarrow\eta_{c}}^{{\rm S}}\left(Q\right)$  & $-0.239$ & $0.154$  & $0.165$\tabularnewline
\hline
\end{tabular}
\par\end{centering}
\protect\caption{Numerical results of relative relativistic corrections $\Delta_{c/g\rightarrow\eta_{c}}\left(Q\right)$.}
%\end{footnotesize}
\label{Tb:rltv_crctn}
\end{table}

At high energy scales, a remarkable increase arises in small $z$ region for the $c\to \eta_c$ fragmentation function when we include
the mixing effect. This can be attributed to the singular behavior of the splitting kernel of $q\to g$ in small $z$.
While the mixing on gluon fragmentation function in small $z$ region is not as oblivious as $c$ quark fragmentation, 
because $g \to q$ and $q\to q$ splitting are all finite in small $z$ region but $g \to g$ splitting is divergent and dominate in
the small $z$ region. Therefore, the singlet evolution is quite important especially for $c$ quark fragmentation case.

\section{Summary}
\label{Summary}

In this work, we have calculated the leading relativistic correction to the
fragmentation function $g\to \eta_c$.
Curiously, the order-$v^2$ fragmentation function shares the identical functional dependence on $z$ as the LO one,
but only differs by an overall factor $-\frac{11}{6} \langle v^2 \rangle$. The relativistic correction appears
to considerably decrease the $g\to \eta_c$ fragmentation probability.
It is interesting to note that, the effect of relativistic correction to this fragmentation function is
 opposite to the radiative correction, which significantly enhances the fragmentation probability for $g\to
 \eta_c$~\cite{Artoisenet:2014lpa}.

We have also confirmed the previous results on the order-$v^2$ correction to the $c\to \eta_c$ fragmentation function.
We further study the evolution the fragmentation functions for $g\to \eta_c$ and $c\to \eta_c$ with respect to the energy scales.
We find that taking into account the mixing in evolution has a striking effect on the small-$z$ behavior of the fragmentation function.

When more copious $\eta_c$ samples are collected at high-$p_T$ in future LHC experiments, it will be
interesting to conduct phenomenological analysis to test our understanding about the $g,c\to \eta_c$ fragmentation mechanism.

\begin{acknowledgments}
We are grateful to Wen-Long Sang for useful discussions. X.-X wishes to thank Alessandro Bacchetta for discussion
on \textsf{HOPPET} package.
The work of X.-R.~G and Y.~J. is supported in part by the National Natural Science Foundation of China under Grants
No.~11475188, No.~11261130311 (CRC110 by DGF and NSFC), by the IHEP Innovation Grant under contract number Y4545170Y2,
and by the State Key Lab for Electronics and Particle Detectors.
The work of L.J.~L. is supported in part by the National Natural Science Foundation of China under Grants
No.~11222549 and No.~11575202.
%-----------------------
\end{acknowledgments}

\clearpage

\end{document}